\documentstyle[prl,aps,preprint,epsf]{revtex}  

\begin{document}

\title{Magnetic Order and Dynamics in an Orbitally Degenerate Ferromagnetic Insulator}
\author{C.~Ulrich$^1$, G.~Khaliullin$^1$, S. Okamoto$^{1,2}$, M. Reehuis$^3$, A. Ivanov$^4$,\\ H. He$^1$,
        Y. Taguchi$^5$, Y.~Tokura$^5$, and B.~Keimer$^1$}
\address{$^1$Max--Planck--Institut f\"ur Festk\"orperforschung, 70569 Stuttgart, Germany}
\address{$^2$ Institute of Physical and Chemical Research (RIKEN),
        Saitama 351-0198, Japan}
\address{$^3$ Hahn--Meitner--Institut, 14109 Berlin, Germany}
\address{$^4$ Institut Laue--Langevin, 156X, 38042 Grenoble, France}
\address{$^5$ Department of Applied Physics, University of Tokyo, 113 Tokyo, Japan}

\date{\today}

\maketitle

\begin{abstract}
Neutron scattering was used to determine the spin structure and
the magnon spectrum of the Mott--Hubbard insulator YTiO$_3$. The
magnetic structure is complex, comprising substantial G-type and
A-type antiferromagnetic components in addition to the predominant
ferromagnetic component. The magnon spectrum, on the other hand,
is gapless and nearly isotropic. We show that these findings are
inconsistent with the orbitally ordered states thus far proposed
for YTiO$_3$ and discuss general implications for a theoretical
description of exchange interactions in orbitally degenerate
systems.
\end{abstract}

\newpage

The interplay between spin and orbital degrees of freedom in
transition metal oxides has been a subject of investigation since
the 1950's. As one of the salient outcomes of this effort, the
``Goodenough-Kanamori rules'' provide a description of the
exchange interactions between magnetic atoms, and hence the
magnetic ordering pattern, in terms of the relative orientation of
valence orbitals on neighboring lattice sites. This field has
recently moved back into the center of attention \cite{tokura}, as
advances in materials preparation have made it possible to
investigate not only the {\it static} spin and orbital
arrangements, but also the spin and orbital {\it dynamics} in
$d$-electron systems. In cubic manganites, the focus of much of
the recent attention, the spin wave excitations have been studied
extensively \cite{lynn}, and excitations associated with the
$d$-orbital degrees of freedom have recently also been detected
\cite{saitoh}.

The cubic crystal field in the perovskite structure splits the
degenerate $d$-orbital manifold of a free transition metal ion
into a lower-lying triplet of $t_{2g}$ symmetry and a higher-lying
$e_g$ doublet. In the manganites, where the $e_g$ doublet is
partially occupied, coupling to the lattice through the
Jahn-Teller effect lifts the orbital degeneracy and generally
pushes the orbital excitations to energies much larger than the
magnon band width. The spin and orbital dynamics are thus largely
decoupled, and quantum effects are suppressed. For $t_{2g}$
orbitals, on the other hand, the higher degeneracy and the more
isotropic, less bond-directional charge distribution reduces the
lattice coupling, and one may expect a more dramatic interplay
between the orbital and the spin dynamics. The manifestations of
this interplay should be most apparent for a single $d$-electron
in the $t_{2g}$ manifold, a situation realized in the pseudocubic
titanates. Indeed, the magnetic properties of the Mott-Hubbard
insulator LaTiO$_3$ are difficult to reconcile with predictions
based on the standard Goodenough-Kanamori rules. The G-type
antiferromagnetic ground state \cite{greedan79} of LaTiO$_3$ is in
conflict with the predictions of all electronic structure
calculations thus far reported \cite{mizokawa,sawada}. Further,
its ordered moment is smaller and its spin wave spectrum is more
isotropic than predicted by conventional superexchange models
\cite{keimer}.

The theoretical implications of the unusual magnetic properties of
LaTiO$_3$ have remained controversial. The theories thus far
proposed invoke disparate effects ranging from a subtle, hitherto
unobserved lattice distortion \cite{imada} to a novel spin-orbital
resonance \cite{Kha00,Kha01}. In order to extend the empirical
basis for these model calculations, we have investigated the
microscopic magnetic properties of YTiO$_3$, a sister compound of
LaTiO$_3$ whose larger O-Ti-O bond angle results in a reduced
electronic band width and an increased Mott-Hubbard gap
\cite{taguchi,okimoto}. In contrast to LaTiO$_3$, the
ferromagnetic ground state of YTiO$_3$ (Ref. \cite{greedan81}) is
in accord with electronic structure calculations
\cite{mizokawa,sawada}. We have used neutron diffraction to show
that the full magnetic structure of YTiO$_3$, while predominantly
ferromagnetic, is actually noncollinear and hence more complex
than previously assumed. Further, an inelastic neutron scattering
study reveals a magnon spectrum that cannot be explained in terms
of the orbital ordering patterns thus far proposed. We discuss our
results in the light of current theories of superexchange in the
presence of orbital degeneracy.

The sample was an untwinned single crystal of volume 0.4 cm$^3$
grown by the floating zone technique as described elsewhere
\cite{taguchi}. The neutron scattering experiments were performed
at the IN22 triple--axis spectrometer at the Institut
Laue--Langevin in Grenoble, France, and at the E5 four--circle
diffractometer at the BER-II reactor of the Hahn--Meitner
Institute in Berlin, Germany. For the measurements on IN22, a
pyrolytic graphite (PG) monochromator and a PG--analyzer, both
horizontally collimating, were used. Depending on kinematic
constraints and resolution requirements, the final wavevector was
fixed at 1.64~\AA$^{-1}$, 2.66~\AA$^{-1}$ or 4.10~\AA$^{-1}$. On
E5, neutron wavelengths 2.36~\AA~ or 0.902~\AA~ were selected by a
PG or a Cu monochromator, respectively, and the data were
collected with a two--dimensional position sensitive
$^3$He--detector.

A set of 1359 structural Bragg reflections taken on E5
\cite{unpublished} could be refined with the orthorhombic space
group $Pnma$ and room temperature lattice parameters $a =
5.3584(9) $~\AA, $b = 5.6956(8)$~\AA~ and $c = 7.6371(11)$~\AA, in
agreement with previous reports \cite{greedan79,zubkov}. In order
to simplify the comparison to model calculations, we sometimes use
the pseudocubic ($c$) unit cell with lattice parameters $a' =
a/\sqrt{2}$, $b'=b/\sqrt{2}$, and $c'=c/2$ instead of the full
orthorhombic ($o$) cell. Our data provide no indication of a
structural phase transition below room temperature.

\begin{figure}[tbp]
\epsfxsize=0.65\hsize \epsfclipon \centerline{\epsffile{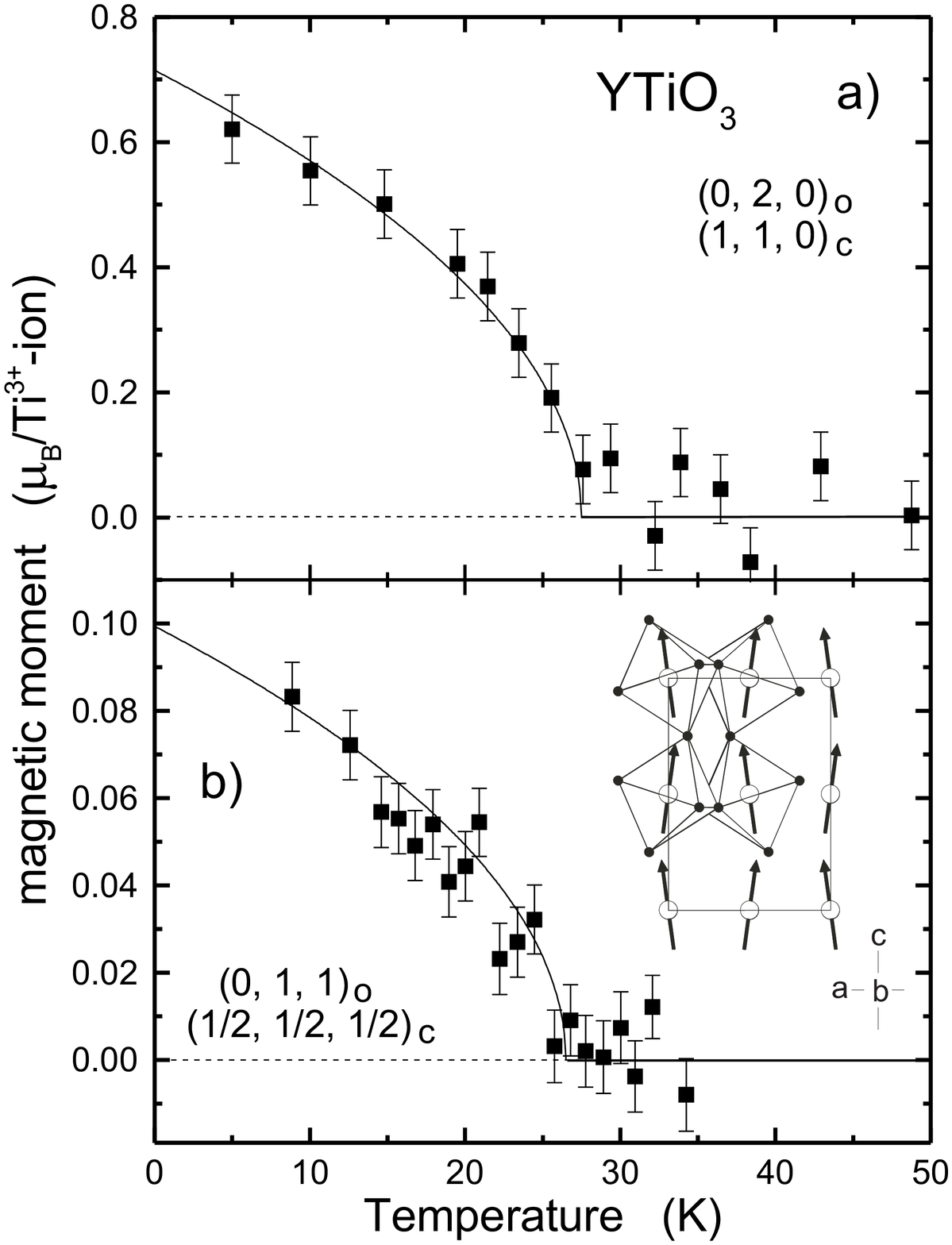}}
\vspace{2ex} \caption{\label{fig1} (a) Ferromagnetic and (b)
G-type antiferromagnetic components of the ordered magnetic moment
of YTiO$_3$ extracted from the amplitudes of the $(0, 2, 0)_o$ and
$(0, 1, 1)_o$ magnetic Bragg reflections, respectively. A weakly
temperature dependent nuclear contribution to both reflections has
been subtracted. The inset gives a pictorial representation of the
magnetic structure. The TiO$_6$ octahedra are highlighted.}
\end{figure}

The presence of magnetic intensity could be established from the
temperature dependence of the primitive Bragg reflections at low
momentum transfer, some of which show a spontaneous increase in
intensity below $T_C = 27$ K (Fig. \ref{fig1}). The reflections
(2, 0, 0)$_o$, (0, 2, 0)$_o$, (1, 1, 0)$_o$ and (1, 1, 2)$_o$ show
the strongest magnetic intensities, indicating a predominantly
ferromagnetic coupling of the Ti-moments as previously reported
\cite{greedan79}. In addition, however, magnetic Bragg reflections
were also observed at (0, 1, 1)$_o$, (1, 0, 1)$_o$, (0, 0, 1)$_o$
and (0, 0, 3)$_o$ (Fig. \ref{fig1}b). This indicates the presence
of hitherto unobserved staggered components of the ordered moment.
A full analysis of the magnetic diffraction pattern at $T=10$~K
gives the following components of the ordered moment per Ti$^{3+}$
ion: a ferromagnetic component of $0.544(11)~\mu_B$ along $c$; a
G-type antiferromagnetic component of $0.082(12)~\mu_B$ along $a$;
and an A-type antiferromagnetic component of $0.047(13)~\mu_B$
along $b$. In agreement with Bertaut's representation analysis
\cite{bertaut}, the magnetic structure can thus be described by
the basis functions $G_x$, $A_y$ and $F_z$ \cite{footnote}. A
pictorial representation is given in the inset to Fig. \ref{fig1}.
The total magnetic moment at 10~K is $0.553(11)~\mu_B / {\rm
Ti}^{3+}$--ion. Since the magnetic Bragg peak intensity is not
saturated at 10~K, this extrapolates to $0.715~\mu_B$ at $T=0$,
somewhat lower than the saturated moment of $0.84~\mu_B$
previously reported for samples with proportionally higher Curie
temperatures \cite{tsubota}.

\begin{figure}[tbp]
\epsfxsize=0.65\hsize \epsfclipon \centerline{\epsffile{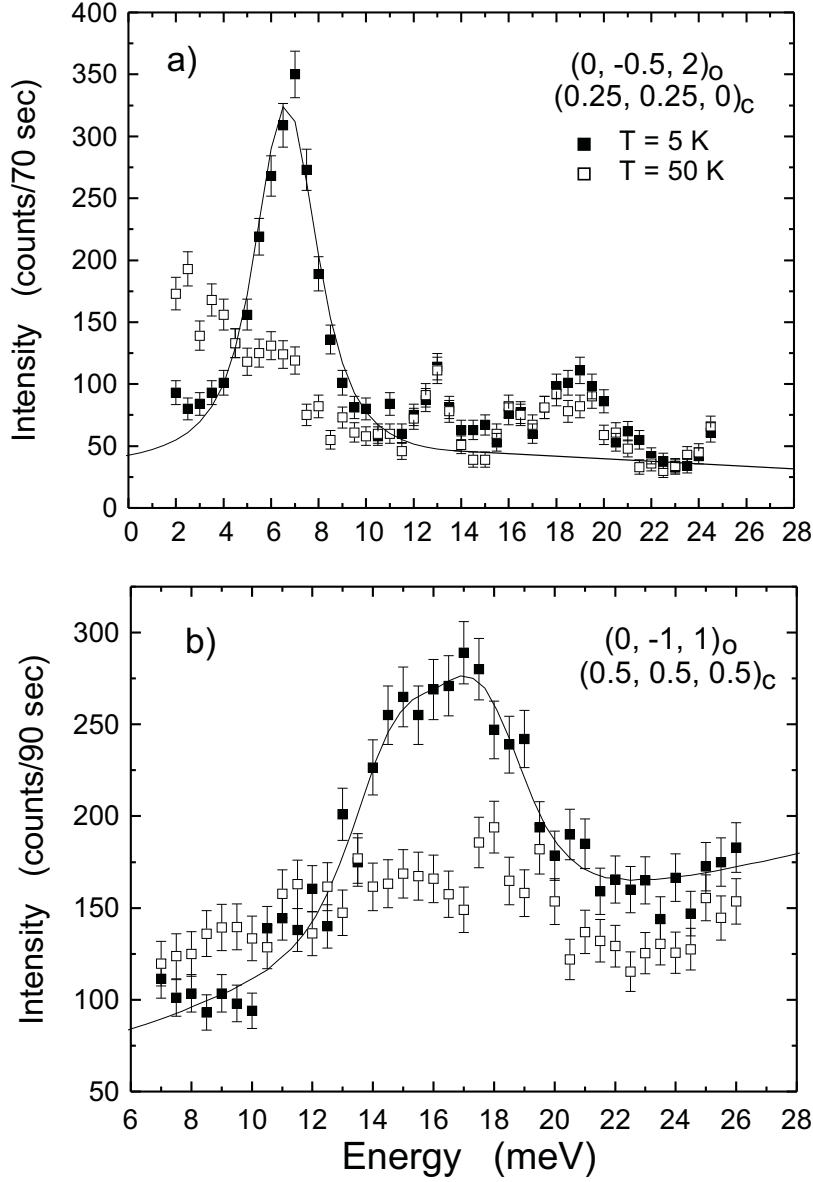}}
\vspace{2ex} \caption{\label{fig2} Representative constant-{\bf q}
scans with (a) ${\bf q}=(\frac{1}{4}, \frac{1}{4}, 0)_c$ and (b)
${\bf q}=(\frac{1}{2}, \frac{1}{2}, \frac{1}{2})_c$ at 5~K
($<T_C$) and 50~K ($>T_C$). At 5 K both phonons and sharp magnons
are present. The magnons become overdamped above $T_C$. The lines
are results of a convolution of a magnon cross section with the
spectrometer resolution function, as discussed in the text.}
\end{figure}

We now turn to the determination of the spin wave spectrum by
inelastic neutron scattering. Fig. \ref{fig2} shows representative
constant--{\bf q} scans at T = 5 K and 50 K, that is, above and
below $T_C$. As shown in the figure, magnons can be distinguished
from phonons on the basis of their temperature dependence. While
the former become overdamped above $T_C$, the latter only show a
slight intensity change due to the thermal population factor.
Since the lineshape of the observed features is strongly
influenced by the spectrometer resolution, a deconvolution is
required to accurately extract the spin wave peak positions. We
have used both the standard Cooper--Nathans procedure and a Monte
Carlo ray--tracing routine \cite{Kulda}. Some of the resulting
profiles are shown in Fig. \ref{fig2}.

\begin{figure}[tbp]
\epsfxsize=0.95\hsize \epsfclipon \centerline{\epsffile{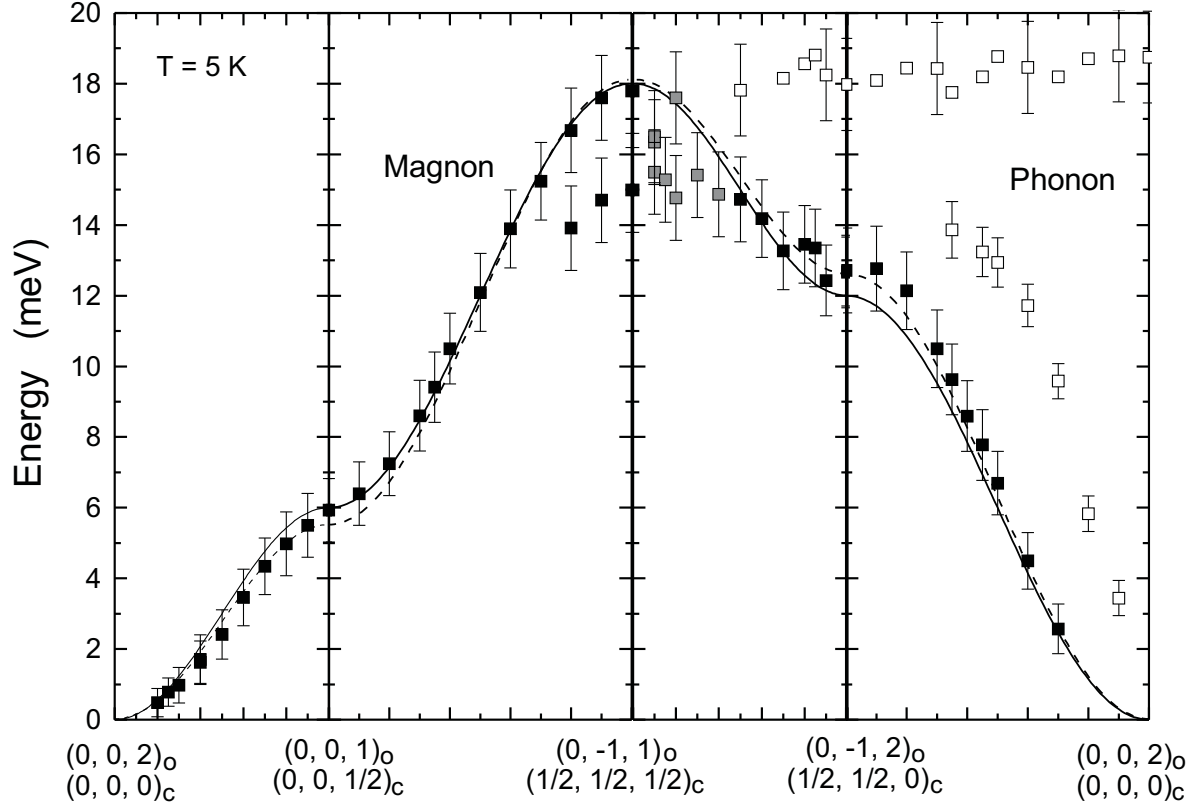}}
\vspace{2ex} \caption{\label{fig3} Spin wave dispersion relations
at $T = 5$ K. The closed symbols refer to magnetic excitations,
whereas the open symbols can be identified as phonons. The grey
points were extracted from scans at low temperatures, without a
corresponding scan above $T_C$, so that an unambiguous
identification is not yet possible. The solid line is the magnon
dispersion relations derived from an isotropic Heisenberg model.
The dashed line is the same model augmented by a spin anisotropy
as discussed in the text.}
\end{figure}

The inelastic neutron scattering data are summarized in Fig.
\ref{fig3}. Some features of the spectrum are immediately
apparent. First, the spectrum is gapless: by high resolution
constant-{\bf q} scans at the zone center we have obtained an
upper bound of 0.3 meV on the spin wave gap. Second, the spectrum
is almost isotropic: a surprisingly good fit can be obtained by an
isotropic Heisenberg model which gives a dispersion of the form
$\hbar \omega = 6 S |J| (1 - \gamma_{\bf q})$ with $\gamma_{\bf
q}= \frac{1}{3} (\cos(q_x a') + \cos(q_y b') + \cos(q_z c'))$ and
$S=1/2$. The best fit of the spectrum is obtained by a
nearest-neighbor superexchange parameter $J=-3.0$ meV,
corresponding \cite{rushbrooke} to $T_C = [1.45 S (S+1) -0.18)]
|J| /k_B = 32$ K, which agrees well with the experimental value. A
small systematic deviation from the result of this fit is
noticeable, albeit within the experimental error bars (Fig.
\ref{fig3}). The fit can be improved by introducing either a
subtle ($\sim 6$\%) difference between the exchange parameters in
the $ab$-plane and along the $c$-axis, or by incorporating a small
exchange anisotropy that is also required to explain the
noncollinear spin structure. Specifically, the leading anisotropy
term in the spin Hamiltonian of the titanates is of cubic symmetry
\cite{Kha01} and modifies the magnon dispersion as follows:

\begin{equation}
\hbar \omega = \sqrt{3 |J| (1-\gamma_{\bf q}) + \Delta +
A(1-\cos(q_x a'))} \times
             \sqrt{3 |J| (1-\gamma_{\bf q}) + \Delta + A(1-\cos(q_y b'))}  \ ,
\end{equation}

where $A$ is the anisotropy parameter and $\Delta = 0.093 A^2 /
|J|$ is the spin wave gap. The best fit gives $A=0.8$ meV and
$J=-2.75$ meV so that $\Delta = 0.02$ meV, consistent with the
experimental upper bound. The cubic anisotropy also induces spin
canting in a pattern identical to the tilting pattern of the
TiO$_6$ octahedra in the $Pnma$ structure. This is indeed the
dominant component of the observed canting pattern (Fig.
\ref{fig1}). An isotropic Heisenberg Hamiltonian with a small
exchange anisotropy thus provides a good description of the
magnetic ground state and dynamics.

While the neutron scattering data can be fitted by single,
resolution limited profiles over most of the Brillouin zone (Fig.
\ref{fig2}a), the profiles near the
$(\frac{1}{2},\frac{1}{2},\frac{1}{2})_c$ point, where the magnon
crosses an optical phonon branch at 19 meV, are broader than the
instrumental resolution (Fig. \ref{fig2}b). The same effective
broadening was also observed in the spin-flip channel of a
polarized-beam experiment. The best overall fit of the
constant-$Q$ scans near the zone boundary is obtained by two-peak
profiles corresponding to two sharp but nondegenerate magnons
(Fig. 2b). The splitting of the magnon degeneracy probably
originates from a hybridization of the magnon with the optical
phonon. Magnon-phonon hybridization has been observed in other
transition metal compounds and was attributed to a modulation of
the crystal field potential by the phonon \cite{lovesey}. However,
as the two peaks are not completely resolved, other origins (such
as a finite magnon lifetime) cannot be ruled out entirely at
present.

Leaving these subtleties aside, we now discuss the overall
features of the magnon spectrum in terms of the orbitally ordered
state predicted by band structure calculations
\cite{mizokawa,sawada} and reported to be consistent with NMR
\cite{itoh} and neutron \cite{akimitsu} form factor data. In this
state, the $d$-electron occupies the following orbitals on the
four inequivalent Ti sites of the $Pnma$ structure
\cite{mizokawa,sawada}:

\begin{eqnarray}
|\psi \rangle _{1,3} & = & c_1 | d_{yz} \rangle \pm c_2 | d_{xy} \rangle , \nonumber \\
|\psi \rangle _{2,4} & = & c_1 | d_{xz} \rangle \pm c_2 | d_{xy}
\rangle. \label{wavefunctions}
\end{eqnarray}

Using the superexchange Hamiltonian given in Ref. \cite{Kha01} and
the above wave functions, the expectation values of the $c$-axis
and in-plane exchange couplings are:

\begin{eqnarray}
J_{c} & = & J_{SE}[(r_1+r_2 r_3)(1-n_{xy}) - (r_1-r_2)](1-n_{xy})/2, \nonumber \\
J_{ab} & = & J_{SE}[(r_1+r_2 r_3) n_{xy}^2 -
(r_1-r_2)(1+n_{xy})/2]/2.
\end{eqnarray}

\begin{figure}[tbp]
\epsfxsize=0.95\hsize \epsfclipon \centerline{\epsffile{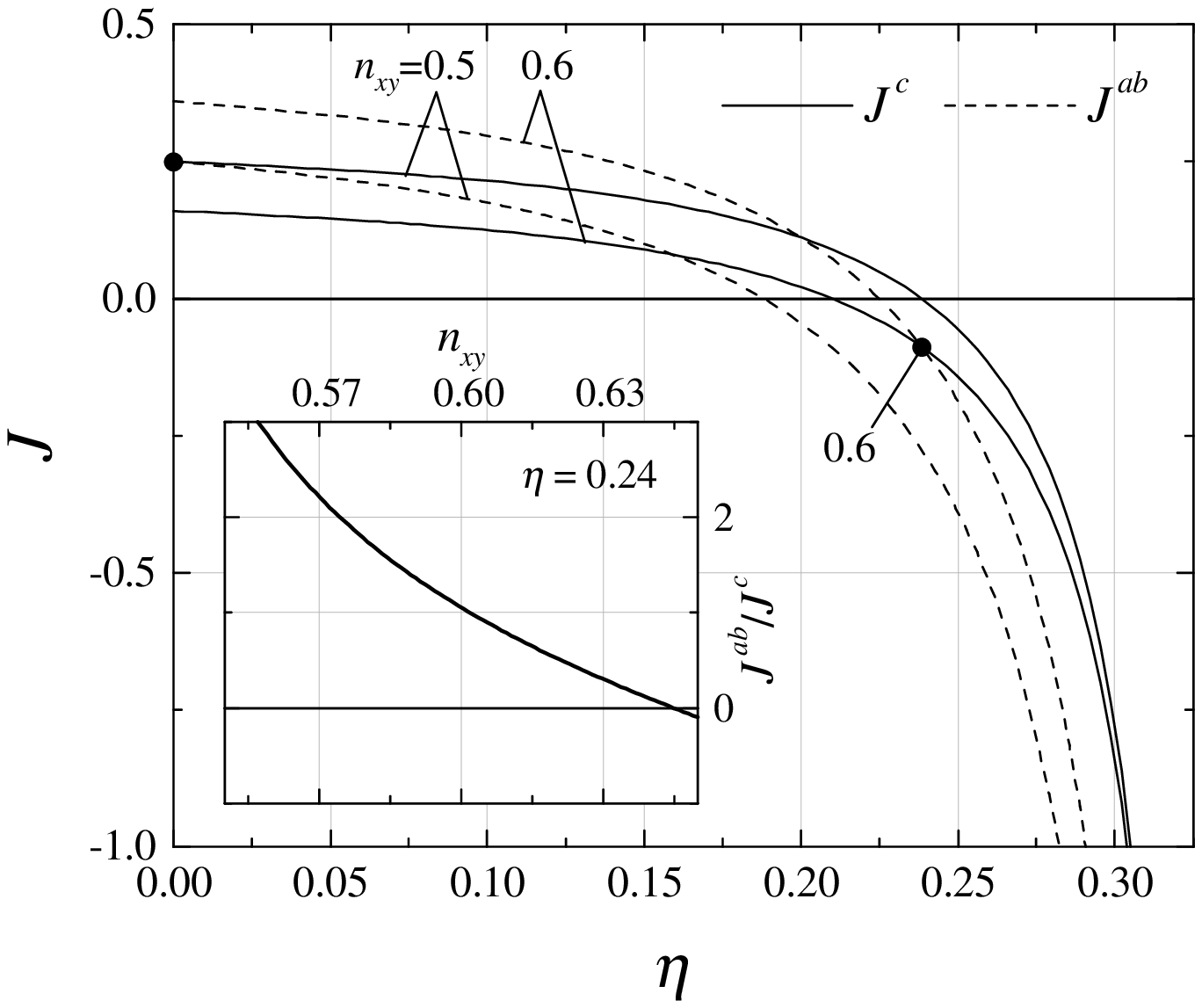}}
\vspace{2ex} \caption{\label{fig4} Exchange constants (in units of
$J_{SE}$) as a function of the Hund coupling $\eta$ for different
values of the parameter $n_{xy}$ characterizing the orbital state.
The inset shows the ratio of in-plane and out-of-plane exchange
couplings for fixed $\eta$ as a function of $n_{xy}$.}
\end{figure}

Here, $J_{SE}=4t^2/U$ (where $t$ stands for the Ti-Ti hopping
parameter and the Kanamori parameter $U$ for the Coulomb repulsion
on the same $t_{2g}$ orbital) represents the overall superexchange
energy scale, and $n_{xy}=c_2^2$ is the fraction of the $d_{xy}$
orbital controlling the orbital state. The coefficients
$r_1=1/(1-3\eta)$, $r_2=1/(1-\eta)$ and $r_3=1/(1+2\eta)$
originate from the Hund's rule splitting of the excited $t_{2g}^2$
multiplet by $\eta=J_H/U$. The parameters $n_{xy}$ and $\eta$
control the exchange couplings along the different axes and their
ratio, as shown in Fig. \ref{fig4}.

In principle, one may obtain isotropic exchange parameters (that
is, $J_c=J_{ab}$) of ferromagnetic sign \cite{ishihara} by using
an orbital state with $n_{xy} \sim 0.6$ (close to the value
reported in Ref. \cite{akimitsu}). However, this requires a large,
unrealistic value of the Hund coupling. For representative values
of $J_H \sim 0.64$ eV and $U - \frac{20}{9} J_H \sim 4$ eV,
\cite{mizokawa} one can estimate $\eta \sim 0.12$ which is
inconsistent with isotropic, ferromagnetic exchange parameters
(Fig. \ref{fig4}). Moreover, an unphysical fine tuning of the
orbital state is required to obtain a spin Hamiltonian of cubic
symmetry. Because of the high sensitivity of the spin interactions
to the orbital state, even a small ($\sim 5$\%) change of $n_{xy}$
leads to a strong spatial anisotropy of the coupling constants and
may even reverse their relative sign (inset in Fig. \ref{fig4}).

An explanation of the small spin wave gap $\Delta \leq 0.3$ meV
presents a further serious difficulty for the conventional
Goodenough-Kanamori picture. The TiO$_6$ octahedra of YTiO$_3$ are
elongated, leaving four almost equal Ti-O bonds
\cite{greedan79,unpublished,zubkov}. To a first approximation, the
crystal-field ground state should therefore be a quadruplet with
substantial unquenched orbital angular momentum and a large magnon
gap. Using the wave functions of Eq. \ref{wavefunctions} with
parameters that result in $J_c = J_{ab} \sim -0.1 J_{SE}$, we have
obtained a quantitative estimate of $\Delta \simeq J_{SE}
(\lambda/D)^2$ for the gap due to the symmetric exchange
anisotropy, where $\lambda \sim 15-20$ meV is the spin-orbit
coupling constant and $D$ is the crystal-field splitting of the
two lowest-lying Kramers doublets. In addition, the
Dzyaloshinskii-Moriya interaction cants the spins by an angle
$\phi \simeq (J_{SE}/3|J|)(\lambda/D)$ away from the $c$-axis and
further increases the gap by $|J| \phi^2$. To be consistent with
the experimentally determined canting angle and upper bound for
the gap, $D$ must be a least 200 meV which is hard to reconcile
with the tetragonal crystal field environment of the Ti$^{3+}$
ion.

These considerations imply the need to go beyond the standard
Goodenough-Kanamori approach in order to arrive at a theoretical
description of the nearly isotropic magnon spectrum of YTiO$_3$.
While the experimental situation \cite{itoh,akimitsu} precludes an
orbitally disordered state such as the one proposed for LaTiO$_3$,
\cite{Kha00,Kha01} quantum zero-point fluctuations in the orbital
sector naturally lead to magnon spectra with diminished spatial
and spin anisotropies even if orbital order is not entirely
obliterated. This has been substantiated in a recent explicit
model calculation \cite{preprint}.

In summary, we have shown that the magnon dispersions of YTiO$_3$
are an extremely sensitive gauge of the orbital state. The
measured dispersions are inconsistent with the orbitally ordered
states thus far proposed and point to the importance of orbital
zero-point fluctuations that go qualitatively beyond the
conventional Goodenough-Kanamori picture. YTiO$_3$ is an excellent
model system for a full quantum many body theory of the
superexchange interaction in orbitally degenerate systems, an
interesting subject of future research.

\newpage

\newpage

\newpage

\begin{references}
\bibitem{footnote} As our data agree very well with an analysis
    based on spin canting alone, we have not considered effects arising from
    a possible alternation of the magnetic form factor due to orbital order,
    as discussed in Ref. \cite{akimitsu}.
\end{references}
\end{document}